\documentclass[11pt]{article}

\usepackage{amsmath}
\usepackage{amsfonts}
\usepackage{amssymb}
\usepackage{graphicx}
\usepackage{pstricks}
\usepackage{pst-plot}
\usepackage[all]{xy}
\usepackage[hang,nooneline]{subfigure}

\begin{document}

\title{Is the physics within the Solar system really understood?}

\author{C. L{\"a}mmerzahl${}^1$, O. Preuss${}^2$, and H. Dittus${}^1$ \\  \\ 
${}^1$ ZARM, University of Bremen, Am Fallturm, 28359 Bremen, Germany \\ 
${}^2$ Max--Planck--Institute for Solar System Research, Max-Planck-Str. 2, \\ 
37191 Katlenburg-Lindau, Germany}


\maketitle

\begin{abstract}
A collection is made of presently unexplained phenomena within our Solar system and in the universe. These phenomena are (i) the Pioneer anomaly, (ii) the flyby anomaly, (iii) the increase of the Astronomical Unit, (iv) the quadrupole and octupole anomaly, and (v) Dark Energy and (vi) Dark Matter. A new data analysis of the complete set of Pioneer data is announced in order to search for systematic effects or to confirm the unexplained acceleration. We also review the mysterious flyby anomaly where the velocities of spacecraft after Earth swing--bys are larger than expected. We emphasize the scientific aspects of this anomaly and propose systematic and continuous observations and studies at the occasion of future flybys. Further anomalies within the Solar system are the increase of the Astronomical Unit and the quadrupole and octupole anomaly. We briefly mention Dark Matter and Dark Energy since in some cases a relation between them and the Solar system anomalies have been speculated.  
\end{abstract}



\section{Introduction}

Progress in physics always has been stimulated by observations which could not been explained within the presently standard physical theories. In the late 19th century the observations and experiments by Bradley as well as Airy who both observed aberration of distant starlight, of Fizeau who observed a dragging of light in moving media not compatible with theory at that time, and finally the experiment of Michelson and Morley who showed that the failure of the application of the non--relativistic mechanically laws to light propagation. All these effects which could not be made compatible with non--relativistic physics without introducing various unnatural elements into the theory, culminated into the invention of Special Relativity. Then, the theoretical incompatibility of Newtonian gravity with Special Relativity as well as the since long observed perihelion shift of Mercury which first has been attributed to systematic errors or solely to be due to the Solar quadrupole moment, lead to the formulation of General Relativity. Later on, the experimental study of atomic spectra which could not be explained using the laws of classical mechanics first led to Bohr's atomic model and, subsequently, to the the various formulations of quantum mechanics. 

The situation of gravitational physics today bears many similarities. At first, the theoretical inconsistency of quantum mechanics and General Relativity makes a new theory combining these two universal theories necessary. Furthermore, there are observations which at least until now and after many years of studies, have not yet found any convincing explanation. These observations are (i) dark energy which is necessary -- under the assumption of the validity of Einstein's equations -- to describe the accelerated expansion of the universe and (ii) dark matter which -- again under the assumption of General Relativity -- is necessary to account for the galactic rotation curves, for observed gravitational lensing of light, and for the structure formation in the early universe. Of a slightly weaker observational basis is (iii) the Pioneer anomaly, an unexplained constant acceleration of the Pioneer 10 and 11 spacecrafts, (iv) the flyby anomaly, an unexplained increase of the velocity of a series of spacecrafts after Earth Gravity Assists, (v) the recently realized increase of the Astronomical Unit defined by the distance of the planets from the Sun by approximately 10 m per century, and (vi) the quadrupole and octupole anomaly which describes the correlation of the low $l$ contributions of the Cosmic Microwave Background to the orientation of the Solar system. 

These six phenomena, including Dark Energy and Dark Matter which at this stage are nothing more than a synonym for these observations, had neither found any convincing interpretation or solution nor culminated into a finally convincing theory. Lacking any explanation until now, these phenomena have the potential to be of importance for a new physics. 

In this paper we describe all these unexplained observations, state the open questions, and suggest new observations and new missions in order to obtain better data for a better analysis of these phenomena.  

\section{Dark matter}

Dark matter has been introduced in order to ''explain'' the gravitational field needed for the galactic rotation curves, the gravitational lensing of galaxies, and the formation of structures in our universe \cite{Sumner02}. It also appears in the spectral decomposition of the cosmic microwave background radiation \cite{Hu01}. Dark matter is needed if one assumes Einsteins field equations to be valid. However, there is no single observational hint at particles which could make up this dark matter. As a consequence, there are attempts to describe the same effects by a modification \cite{Sanders84} of the gravitational field equations, e.g. of Yukawa form, or by a modification of the dynamics of particles, like the MOND ansatz \cite{Milgrom02,SandersMcGough02}\footnote{In a nice short paper M. Veltman \cite{Veltman03} speculates how astronomers may build up laws of gravity by observing gravity on larger scales (scale of galaxies and of the universe) and compares that with a quantum field theory approach.}, recently formulated in a relativistic frame \cite{Bekenstein04}. Due to the lack of direct detection of Dark Matter particles, all those attempts are on the same footing. 

\section{Dark energy}

Similarly, recent observations of type Ia supernovae indicate that the expansion of the universe is accelerating and that 75\% of the total energy density consist of a dark energy component with negative pressure \cite{Riessetal98,Perlmutteretal99}. Furthermore WMAP measurements of the cosmic microwave background \cite{Bennetetal03,Spergeletal03}, the galaxy power spectrum and the Lyman-alpha forest data lines \cite{BruckPriester98,OverduinPriester01,Tegmarketal04} also indicate -- when compared with standard cosmological models -- the existence of the mysterious Dark Energy that leads to the acceleration of the universe, rather than a modification of the basic laws of gravitation \cite{PeeblesRatra03}. However, also in this case there are attempts to give an explanation in terms of modified field equations, see, e.g., \cite{NojiriOdintsov06}. Recently it has been claimed that dark energy or, equivalently, the observed acceleration of the universe can be explained by inhomogeneous cosmological models, such as the spherically--symmetric Lemaitre--Tolman--Bondi model, see, e.g., \cite{Celerier00,VanderveldFlanaganWasserman06,Apostolopoulosetal06}.

\section{The Pioneer anomaly} 

The Pioneer anomaly is an anomalous unexplained acceleration of the Pioneer 10 and 11 spacecraft of 
\begin{equation}
a_{\rm Pioneer} = \left(8.74 \pm 1.33\right) \cdot 10^{-10}\;{\rm m/s^2} \label{PioneerAcceleration}
\end{equation}
toward the Sun \cite{Andersonetal98,Andersonetal02}. This acceleration seems to have been turned on after the last flyby at Saturn and stayed constant within an 3\% range. 

\subsection{The observation}

The principle of observation was 2--way Doppler tracking: A sender on the Earth emits a signal of frequency $\nu_0$ which is ''seen'' by the spacecraft as frequency 
\begin{equation}
\nu^\prime = \frac{1}{\sqrt{1 - v^2/c^2}} \left(1 - \frac{v}{c}\right) \nu_0 \, .
\end{equation}
The spacecraft sends this frequency back (with a slight offset what, however, will not affect the principle of measurement), so that the receiver on Earth observes the frequency
\begin{equation}
\nu^{\prime\prime} = \frac{1}{\sqrt{1 - v^2/c^2}} \left(1 - \frac{v}{c}\right) \nu^\prime \, .
\end{equation}
The comparison of the sent and received frequency gives the velocity of the spacecraft
\begin{equation}
y = \frac{\nu^{\prime\prime} - \nu_0}{\nu_0} = - 2 \frac{v/c}{1 + v/c} \approx - 2 \frac{v}{c} \, .
\end{equation}
This measured frequency can be compared with the frequency obtained from the calculated orbit given by the gravitational field inside the Solar system together with all kinds of modeling needed (see below). 

The outcome the observation was a continuous drift between the observed and calculated frequency shift  
\begin{equation}
\frac{d (y_{\rm obs} - y_{\rm calc})}{dt} = \left(3.84 \pm 0.01\right) \cdot 10^{-18}\;{\rm s}^{-1} \, .
\end{equation}
This corresponds to a continuous drift in the velocity of the spacecraft or, equivalently, in a constant acceleration (\ref{PioneerAcceleration}).

\subsection{Orbit determination}

Since the observations are made with tracking stations on the moving Earth observing the frequency of signals, the orbit determination consists of five segments:
$$
\left.\hbox{\begin{minipage}{8cm}
\begin{itemize}\itemsep=-2pt
\item Model of gravitational forces
\item Model of external non--gravitational forces
\item Model of internal (spacecraft) non--\\ gravitational forces
\item Model of observation stations
\item Model of signal propagation
\item Codes
\end{itemize}
\end{minipage}} \right\} \;\; \Longrightarrow \;\; \hbox{Orbit and velocity determination} 
$$
We just mention the main aspects of this scheme. Most of this can be found in \cite{Andersonetal02}.

\begin{description}
\item[Gravitational forces] The calculation of the orbits have been performed using relativistic equations of motion for celestial bodies including order $v^4/c^4$
\begin{itemize}\itemsep=-2pt
\item The relativistic gravitational accelerations (EIH model) include the Sun, the Moon, and the 9 planets as point masses in an isotropic PPN $N$--body metric.
\item Newtonian gravity from large asteroids is included. Furthermore, 
\item Terrestrial and lunar figure effects,
\item Earth tides,
\item and Lunar physical librations have been considered.
\end{itemize}
\item[External non--gravitational forces] These forces include:
\begin{itemize}\itemsep=-2pt
\item Solar radiation and Solar wind pressure.
\item Drag from interplanetary dust.
\end{itemize}
\item[Internal (spacecraft) non--gravitational forces] These forces include:
\begin{itemize}\itemsep=-2pt
\item Thermal radiation.
\item Attitude--control propulsive maneuvers and propellant (gas) leakage from the spacecraft's propulsion system.
\item Torques produced from the above forces.
\end{itemize}
\item[Model of observation stations]
An orbit determination has to include a model of the ground stations. This is based on:
\begin{itemize}\itemsep=-2pt
\item Precession, nutation, sidereal rotation, polar motion, tidal effects, and tectonic plates drift. All the informations on tidal deceleration, non--uniformity of rotation, Love numbers, and Chandler wobble have been obtained from LLR, SLR, and VLBI (from ICRF) measurements.
\item Model of DSN antennae and their influence on the tracking data.
\end{itemize}
\item[Modeling of signal propagation] The propagation of the radio signals includes
\begin{itemize}\itemsep=-2pt
\item a relativistic model for light propagation including order $v^2/c^2$.
\item and dispersion due to Solar wind and interplanetary dust.
\end{itemize}
\item[Codes]
Four independent codes have been used for the orbit determination:
\begin{itemize}\itemsep=-2pt
\item JPL Orbit Determination Program (various generations from 1970 -- 2001).
\item The Aerospace Corporation code POEAS (during period 1995 -- 2001).
\item Goddard Space Flight Center conducted a study in 2003 (data from NSSDC).
\item Code of University of Oslo.
\end{itemize}
\end{description}

The definition of these models have to be complemented by a discussion of possible errors. This is tantamount to the search for possible conventional explanations of the effect. We present a few points only. 

\subsection{Discussion of some conventional effects}

In the following we discuss recent and ongoing work on conventional effects which may contribute to errors or perhaps also may be responsible for the observed acceleration. Not included here is the spin--rotation coupling \cite{AndersonMashhoon03}

\paragraph{Dust}

The interplanetary medium consists of (i) interplanetary dust and of (ii) interstellar dust. The first one consists of hot wind plasma (mainly protons and electrons) distributed within the Kuiper belt (from 30 AU to 100 AU). The density of this plasma has been modeled to be of the order $\rho_{\rm IPD} \leq 10^{-24}\;{\rm g/cm^3}$ (Man and Kimura 2000). The interstellar dust which can be distinguished from the interplanetary dust by its greater impact velocity has been measured by Ulysses to have a density of $\rho_{\rm ISD} \leq 3 \cdot 10^{-26}\;{\rm g/cm^3}$. 

The drag acceleration of a spacecraft moving through dust of density $\rho$ is given by 
\begin{equation}
a_{\rm drag} = - K_{\rm s} \rho v_{\rm s}^2 \frac{A_{\rm s}}{m_{\rm s}} \, , \label{DragForce}
\end{equation}
where $K_{\rm s}$ is the satellite's drag coefficient which can be taken to be $\approx 2$. If we assume the drag acceleration to be the observed anomalous acceleration of the Pioneer spacecraft, then this needs a density which is $3 \cdot 10^5$ larger than the interplanetary dust. Therefore dust cannot be the origin of the Pioneer acceleration \cite{NietoTuryshevAnderson05}.

\paragraph{Additional masses in the Solar system}

Additional masses may not only be present in the form of dust but also in form of larger particles. Irrespective of being dust and of the size of these particles, any additional mass will act as an additional gravitational field which may decelerate the spacecraft when leaving the Solar system. Nieto \cite{Nieto05} has calculated analytically the gravitational effect of various configurations, that is, shells, thin rings and wedges of various density profiles. He obtained that for rings with a density falling off with $1/r$ as well as a wedge with a density falling off like $1/r^2$ yields a nearly constant acceleration (neglecting discontinuities at the sharp boundaries of the matter distributions which are, of course, just results of the mathematical model). However, in order the constant acceleration to be of the order of the observed Pioneer acceleration, the mass of the thin ring or the wedge has to be of about 100 time the mass of the Earth which is, by far, not compatible with the observations of, e.g., comets. 

\paragraph{Accelerated Sun}

A nongravitational acceleration of the Sun orthogonal to the ecliptic will also cause an acceleration toward the Sun. Such an accelerated Sun is the consequence of an exact solution of the Einstein equation, the so called $C$--metric \cite{BiniCherubiniMashhoon04}. In the frame of an accelerated Sun, the equation of motion for test masses reads
\begin{equation}
\ddot{\mbox{\boldmath$r$}} + G M_\odot \frac{\mbox{\boldmath$r$}}{r^3} + {\mbox{\boldmath$a$}}_\odot = 0 \, ,
\end{equation}
where $\mbox{\boldmath$r$}$ is the distance between the Sun and the test mass. This gives a constant acceleration toward the Sun \cite{BiniCherubiniMashhoon04}. However, in order obtain an acceleration of the order of the Pioneer anomaly, the acceleration of the Sun orthogonal to the ecliptic has to be larger than what would be obtained if all radiation of the Sun is emitted in one direction. 

\paragraph{Cosmic expansion}

Due to the quite good equality $a_{\rm Pioneer} \approx c H$ where $H$ is the Hubble constant, it has been speculated whether the cosmic expansion has some influence on the (i) signal propagation, (ii) trajectory of the spacecraft, (iii) the magnitude of the gravitational field inside the Solar system, or on (iv) the definition of the distance, that is, the definition of the Astronomical Unit. 


The influence of the expansion of the universe on the procedure of Doppler tracking is negligible. For an expansion described by an Einstein--de Sitter universe
\begin{equation}
ds^2 = - dt^2 + R^2(t) \left(dx^2 + dy^2 + dz^2\right)
\end{equation}
we obtain the conserved quantities
$
\nu_u(t) R(t) = const. \, , 
$ 
where $\nu = \nu_u = k(u)$ is the measured frequency of a light ray $g(k, k) = 0$; $u$ with $g(u,u) = 1$ is the 4--velocity of an observer at rest in the cosmic substrate. To first order in the expansion,
\begin{equation}
\nu_u(t) = \frac{R(t_0)}{R(t)} \nu_u(t_0) \approx \left(1 - H (t - t_0)\right)  \nu_u(t_0)\, , \label{CosmRedShift}
\end{equation}
where $H = \dot R/R$ is the Hubble constant. This describes the Hubble red shift. 

For massive particles we have the conserved quantity 
\begin{equation}
R^2(t) \frac{dr(s)}{ds} = const. \quad \Leftrightarrow \quad R(t) \frac{1}{\sqrt{1 - V^2(t)}} V(t) = const.  \, ,
\end{equation}
where $V$, defined by $g(u,v) = 1/\sqrt{1 - V^2/c^2}$, is the measured velocity of an object moving with 4--velocity $v$ along a geodesic, $D_v v = 0$. 
For small velocities $R(t) V(t) = const.$ implying a slowing down of the velocity
\begin{equation}
V(t_2) = \frac{R(t_1)}{R(t_2)} V(t_1) = \left(1 - H (t_2 - t_1)\right) V(t_1) \label{SlowingDown}
\end{equation}

A distance $D$ measured by time--of--flight of light rays is defined by 
$
D = R (r_2 - r_1) \, .
$ 
The measured velocity of a moving object is then
\begin{equation}
\frac{d}{dt} D = \dot R (r_2 - r_1) + R \dot r_2 = H D + V_2 
\end{equation}
where $V_2$ is the velocity of the object measured with respect to the cosmological substrate. As a consequence, the trajectory of an object which has constant distance to an observer, $0 = \dot D =  H D + V_2$, has to move with velocity 
\begin{equation}
V_2 = - H D \label{VelUnitDistance}
\end{equation}
with respect to the substrate. 

Therefore, the two way Doppler ranging is influenced by three effects: (i) the cosmological redshift (\ref{CosmRedShift}), (ii) the slowing down of the velocity of the spacecraft (\ref{SlowingDown}), and the velocity of the unit distances with respect to the cosmological substrate (\ref{VelUnitDistance}). Taking all together results in the final two-way Doppler ranging signal
\begin{equation}
\frac{\Delta \nu}{\nu_0} = - 2 V(t_1) (1 - H (t_2 - t_1)) \, ,
\end{equation} 
where $t_1$ and $t_2$ are the cosmological time parameters for the emission and reception of the signal. The expansion induced effect is a chirp of Doppler signal related to an acceleration
\begin{equation}
a = H V = \frac{V}{c} c H 
\end{equation} 
which is by a factor $V/c$ smaller than observed Pioneer acceleration $a_{\rm Pioneer} \sim c H$. Therefore, Doppler tracking in an expanding universe cannot account for the observed Pioneer Anomaly.

Furthermore, the influence of the expansion of the universe on the gravitational field of the Sun or the planetary orbits is much too small to be of any influence, see Sec.\ref{GenCosmExp}. 

\paragraph{Drift of clocks on Earth}

Though a drift of clocks by itself is non--conventional physics (the drift of atomic clocks with respect to, in our case, gravitational time scales needs some ''new physics'') it may yield a conventional explanation of the Pioneer anomaly. A quadratic drift of the timekeeping of clocks on Earth may simulate the Pioneer anomaly \cite{Andersonetal02,Ranada05}
\begin{equation}
t \rightarrow t + \frac{1}{2 c} a_{\rm P} t^2 \, . \label{PioneerClockDrift}
\end{equation}
The numerical value is $\frac{1}{2 c} a_{\rm P} \approx 10^{-18}\;{\rm s}^{-1}$. If we assume such a kind of clock drift then the question arises whether this clock drift is consistent with the measurements from other satellites and, in particular, with the ranging of satellites. Another question is whether this time drift is also consistent with the observations of pulsars and binary systems which also define clocks. Pulsars are very stable clocks \cite{Wex01}. Owing to the radiation of gravitational waves the revolution time of binary systems goes down. However, also in this case the stability of this process can be defined \cite{MatsakisTaylorEubanks97}. A comparison with a drift of Earth clocks has not been carried through. 

\medskip
To sum up, the expansion of our universe seems to be of no relevance for the occurrence of the Pioneer acceleration. 

\subsection{Outlook}

\begin{table}
\caption{Data sets analyzed previously and to be analyzed in the future. \label{DataSetTable}}
\begin{center}
\begin{tabular}{|l|ll|ll|} \hline
& \multicolumn{2}{|c|}{previously used data set} & \multicolumn{2}{|c|}{data set to be analyzed} \\
& time span AU & distance & time span AU & distance \\ \hline 
Pioneer 10 & 3.1.1987 -- 22.7.1998 & 40 -- 70.5 & 8.9.1973 -- 27.4.2002 & 4.6 -- 80.2\\
Pioneer 11 & 5.1.1987 -- 1.10.1990 & 22.4 -- 31.7 & 10.4.1974 -- 11.10.1994 & 1.0 -- 41.7 \\ \hline
\end{tabular}
\end{center}
\end{table}

New activities are planned for the very near future. During the last months the complete set of Pioneer data have been recovered \cite{Turyshevetal06,TothTuryshev06} and brought into a digital form readable by modern computers. These data are now ready for a new data analysis covering all data and the total mission duration, see table \ref{DataSetTable}. This new data analysis will be carried through at ZARM and at JPL. It is important to find out, e.g., whether the anomalous acceleration was really not present before the last flyby. Furthermore, a new Deep Space Gravity Explorer mission has been proposed \cite{Dittusetal05}.

\section{The flyby anomaly}

\subsection{The observations}

\begin{figure}[bh!]
\begin{center}
\subfigure[][Two--way S--band Doppler residuals and range residuals during the first Galileo flyby]{\includegraphics[scale=3.45]{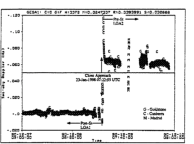} \quad
\includegraphics[scale=3.44]{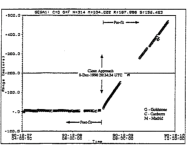}}

\subfigure[][Two--way X--band Doppler residuals and range residuals during NEAR flyby]{\includegraphics[scale=3.5]{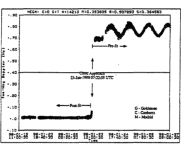} \quad 
\includegraphics[scale=3.5]{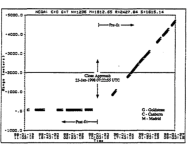}}
\end{center}
\caption{Galileo and NEAR flyby data (from \cite{AntreasianGuinn98}). \label{Fig:FlybyData}}
\end{figure}

It has been observed at various occasions that satellites after an Earth swing--by possess a significant unexplained velocity increase by a few mm/s. This unexpected and unexplained velocity increase is called the {\it flyby anomaly}. According to information from \cite{AndersonWilliams01,AntreasianGuinn98,Morley05} the observed flybys are listed in Table \ref{Table:flyby}. For the actual data for the Galileo and NEAR flyby see Fig.\ref{Fig:FlybyData}.

\begin{table}[h!]
\caption{Observed flybys. \label{Table:flyby}}
\begin{center}
\begin{tabular}{|llcccc|}\hline
Mission & agency & year & pericentre & eccentricity & velocity increase \\ \hline 
Galileo & NASA & Dec 1990 & 959.9 km & 2.47 & $3.92 \pm 0.08$ mm/sec \\
Galileo & NASA & Dec 1992 & 303.1 km & 2.32 & no reliable data${}^{\rm a}$  \\ 
NEAR & NASA & Jan 1998 & 538.8 km & 1.81 & $13.46 \pm 0.13$ mm/sec \\
Cassini & NASA & Aug 1999 & 1173 km & 5.8 & 0.11 mm/sec \\ 
Stardust & NASA & Jan 2001 & 5950 km & & no reliable data${}^{\rm b}$ \\
Rosetta & ESA & Mar 2005 & 1954 km & 1.327 & $1.82 \pm 0.05$ mm/s \\ 
Hayabusa & Japan & May 2004 & 3725 km & ?? & no data available \\
MESSENGER & private & Aug 2005 & ?? & ?? & no data available${}^{\rm c}$ \\ \hline
\end{tabular}
\end{center}
${}^{\rm a}$ too low orbit with too large atmospheric drag 

${}^{\rm b}$ thruster activities

${}^{\rm c}$ US spacecraft operated by a private company 
\end{table}

\begin{figure}[h]
\psset{unit=0.7cm}
\begin{center}
{\footnotesize\begin{pspicture}(-9,-1)(9,8)
\rput(-8.3,0){
\psaxes[Dy=2,dy=1]{->}(7,7)
\rput(7,-0.5){$e$}
\rput(-1,7){\txt{$\Delta v$ \\ [mm/s]}}
\psline[linestyle=dashed](1,0)(1,7)
\psline[showpoints=true](1.327,0.91)(1.81,6.73)(2.47,1.96)(5.8,0.06)
\rput(1.9,0.5){Rosetta}
\rput(1.81,7.1){NEAR}
\rput(3.47,2.3){Galileo}
\rput(5.8,1){Cassini}
}
\rput(0.7,0){
\psaxes[Dy=2,dy=1,Dx=250,dx=1]{->}(8,7)
\rput(8.2,-0.8){\txt{$r_{\rm p}$ \\ [km]}}
\rput(-1,7){\txt{$\Delta v$ \\ [mm/s]}}
\psline[showpoints=true](2.1552,6.73)(3.8396,1.96)(4.692,0.06)(7.816,0.91)
\rput(7.816,1.3){Rosetta}
\rput(2.1552,7.1){NEAR}
\rput(3,1.9){Galileo}
\rput(5.2,0.9){Cassini}
}
\end{pspicture}}
\end{center}
\caption{The velocity increase $\Delta v$ as function of the eccentricity and of the perigee. \label{DeltaVFunction}}
\end{figure}
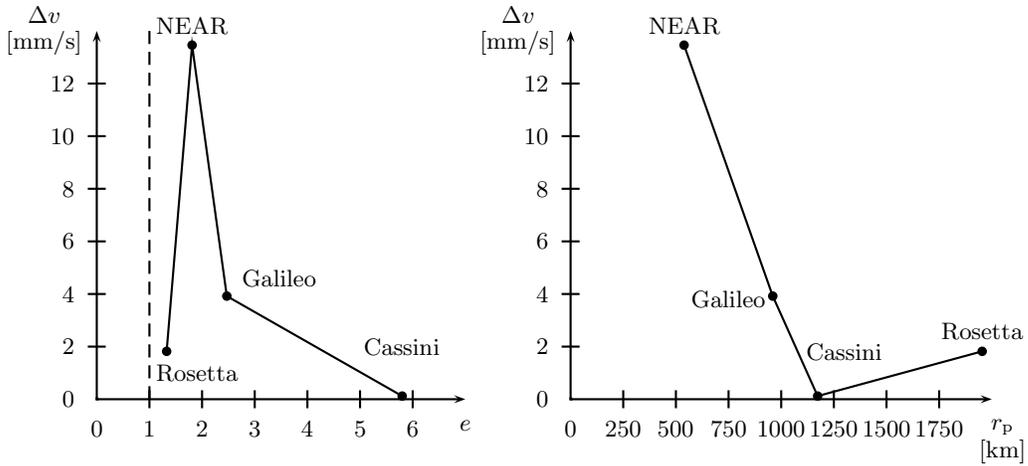

The data can be put into diagrams where the velocity increase can be plotted as a  function of the two orbit parameters eccentricity $e$ and pericentre $r_{\rm p}$, see Fig.\ref{DeltaVFunction}. In general, this is a three--dimensional plot $\Delta v = f(e, r_{\rm p})$. For a plot of this surface four data points are far too less. Therefore, in Fig.\ref{DeltaVFunction} a plot of the velocity increase $\Delta v$ as function of $e$ and of $r_{\rm p}$, respectively, is given. Though from four data points it is much too early to draw any serious conclusion one may speculate the following: If the velocity increase really is due to an unknown gravitational interaction, then it makes sense that (i) the effect goes down with increasing eccentricity, since for larger eccentricity the strength and duration of the interaction with the gravitational field of the Earth goes down, and (ii) that it also should go down for an eccentricity approaching $e = 1$ because the transition to bound orbits, where no effect has been seen, certainly should show no discontinuity due to many kinds of disturbances like drag and non--ideal circumstances like gravitational multipoles, etc. But relying on this poor data base this interpretation is pure speculation only. 

The main problem is not just the limited number of flybys for which sufficiently precise data are publicly available so that the anomaly can be seen at all. Even these available data suffer from low cadence (the anomaly often appears between two data points) and so far only allow an anomaly in the speed, but not in the direction of motion etc. to be identified. Precise data at a much higher cadence of all the motion
parameters of the spacecraft prior to, during and after the flyby would allow a qualitatively improved analysis. 

\subsection{Error analysis}\label{flybyaccel}

As a first remark we give the order of the acceleration leading to the velocity increase. This anomalous acceleration, estimated by the velocity increase during the time of flight near the Earth, is of the order $10^{-4}\;{\rm m/s^2}$. This is considerably larger than the above discussed Pioneer anomaly. Below we will use this acceleration as an approximate value to be compared with disturbing influences. 

It should be kept in mind that the velocity increase has been observed in the two--way Doppler measurements as well as in the ranging measurements. 

Before one starts with possible fundamental explanations of this effect, a serious and reliable error analysis has to be carried through. This analysis has to cover (i) atmospheric mismodeling, (ii) ocean tides, (iii) if the spacecraft becomes charged, then it may experience an additional force due to the Earth's magnetic field, (iv) also the interaction of a hypothetical magnetic moment of the spacecraft with the Earth's magnetic field may give an additional force, (v) ion plasma drag, (vi) Earth albedo, and (vii) Solar wind. Here we give very short first order estimates on these various effects which certainly has to be improved, and show that even with very rough pessimistic assumptions none of these can be held responsible for the flyby anomaly.

\begin{figure}[t]
\begin{center}
\includegraphics[scale=0.5]{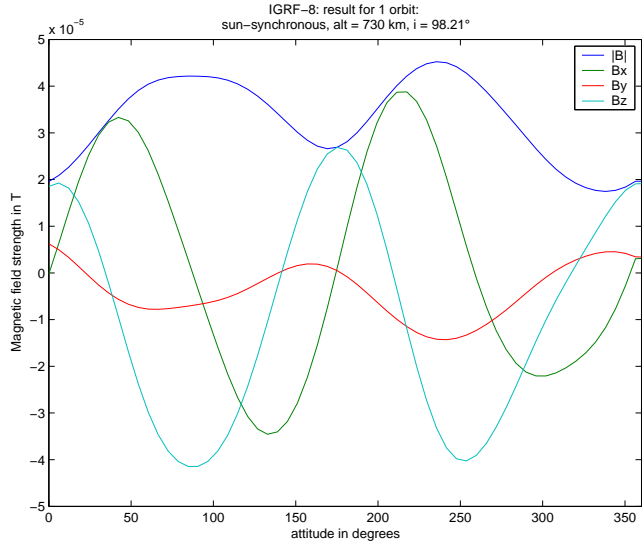}
\end{center}
\caption{The magnetic field of the Earth as function of attitude. \label{MagneticField}}
\end{figure}

\begin{description}
\item[Atmosphere] If a spacecraft of mass $m_{\rm s}$ and effective area $A_{\rm s}$  moves with velocity $v_{\rm s}$ through a medium of density $\rho$, then it experiences a drag acceleration again given by (\ref{DragForce}). 
For a mass of 1 t, an area of $2\;{\rm m}^2$, a velocity of 30 km/s and an atmospheric density at 1000 km height of approx $\rho \approx 10^{-14}\;{\rm kg/m}^3$ we get an acceleration of $a_{\rm drag} \approx 4 \cdot 10^{-8}\;{\rm m/s}^2$ what is far too small to be of any relevance for our problem. Furthermore, this acceleration due to drag has the wrong sign. 
\item[Ocean tides] The ocean tides will lead to a change of the Earth's surface of the order of $\delta r \pm 10\;{\rm m}$. This means that the corresponding quadrupole part of the Earth's gravitational potential is of the order $\epsilon = 2 \delta r/R_\oplus$ smaller than the monopole part of the Earth, where $R_\oplus$ is the radius of the Earth. Since $\epsilon \approx 10^{-6}$, the corresponding additional acceleration also is factor $10^{-6}$ smaller than the ordinary acceleration from the monopole part of the Earth's gravitational field. The latter being less than $10 \; {\rm m/s}^2$, the acceleration due to tides is at most $10^{-5}\;{\rm m/s}^2$ and, thus, cannot be responsible for the flyby anomaly. 
\item[Solid Earth tides] Since Earth solid tides are much smaller than ocean tides, the analysis above shows that this cannot cause the effect. 
\item[Charging of the spacecraft] In a recent study of charging of the LISA test masses \cite{Sumneretal04} the charging has been estimated by $10^{-10}\;{\rm C}$. So, for the whole satellite it might be a conservative assumption that the charge is less than $Q \leq 10^{-7}\;{\rm C}$. A satellite of 1 t carrying a charge $Q$ and moving with $v = 30\;{\rm km/s}$ in the magnetic field of the Earth which is of the order 0.2 G will experience an acceleration $10^{-8}\;{\rm m/s}^2$ far below the observed effect. 
\item[Magnetic moment] The force on such a body carrying a magnetic moment $\mbox{\boldmath$m$}$ moving in a magnetic field $\mbox{\boldmath$B$}$ is $\mbox{\boldmath$F$} = \mbox{\boldmath$\nabla$} (\mbox{\boldmath$m$} \cdot \mbox{\boldmath$B$})$. Since the magnetic moment of a spacecraft is not more than $2\;{\rm A}\;{\rm m}^2$ and the steepness of the magnetic field can be estimated by $|\Delta B/\Delta x| \leq 2 \cdot 10^{-7} \;{\rm G/m}$, see Fig.\ref{MagneticField}, the maximum force of a spacecraft is $F \leq 4 \cdot 10^{-11} \;{\rm N}$ implying typically a maximum acceleration of $4 \cdot 10^{-15} \;{\rm m/s}^2$ which safely can be neglected. 
\item[Earth albedo] The Earth albedo causes a pressure on the spacecraft of approx  $1\;\mu{\rm N}/{\rm m}^2$ which leads, for an effective area of $2\;{\rm m}^2$ to a force of $2.4\;\mu{\rm N}$. For a mass of the spacecraft of 1 t this will give an acceleration of $a_{\rm albedo} \approx 2.4 \cdot 10^{-9} \; {\rm m/s}^2$ what can be neglected compared to the searched for effect of $10^{-4}\;{\rm m/s}^2$. 
\item[Solar wind] The solar wind exerts on spacecraft a pressure of approx $4\;\mu{\rm N}/{\rm m}^2$ which gives an acceleration of max $a_{\rm solar\; wind} \approx 2.4 \cdot 10^{-9} \; {\rm m/s}^2$ which again can be safely neglected.
\item[Spin--rotation coupling] A coupling of the helicity of the radio waves with the rotation of the spacecraft and the rotation of the Earth also leads to an effect which simulates a changing velocity \cite{AndersonMashhoon03}. This, however, applies to the two--way Doppler data only. Since simultaneously also ranging, what is independent of the helicity--rotation coupling, indicated an increase of the velocity, spin--rotation cannot be responsible for this observation. 
\end{description}
Also estimates of the influence of the Moon including Moon oblateness, the Sun, other planets, relativistic effects, and indirect oblateness of the Earth have been shown to be order of magnitude smaller than the observed effect  \cite{AntreasianGuinn98}.

None of these disturbing effects could explain the flyby--anomaly. 

\subsection{Explanations from ''new physics''} 

As reported in the paper \cite{AntreasianGuinn98} several non--standard physical models have been used to explain this velocity increase. Since these considerations have not been published we just mention the discussed models:
\begin{itemize}\itemsep=-2pt
\item Non--conservative potential energy.
\item Non--Newtonian gravity (e.g. Moffat, Yukawa, etc).
\item PPN.
\item Modifications of relativity.
\item Torsion, i.e. the {\it eps2 model}. This model is said to fit the data, but is not compatible with stability of planetary orbits. 
\end{itemize}
As told in \cite{AntreasianGuinn98} none of these could explain the flyby--anomaly. Being so large and so near to the Earth, the expansion of the universe should play no role for this anomaly. In Sec.\ref{Description} we sketch a general approach to a description of the motion of test bodies. Among the terms found there are several which may be considered as phenomenological description of a velocity increase. In order to be able to pin down a specific interaction term it is necessary to make detailed studies of the nature of the velocity increase:

\subsection{Future flybys} 

In the near future there will be two flybys, both by Rosetta \cite{Morley05}
\begin{itemize}\itemsep=-2pt
\item Rosetta: flyby on 13 November 2007 (pericentre altitude 4942 km).
\item Rosetta: flyby on 13 November 2009 (pericentre altitude 2483 km).
\end{itemize}
We strongly suggest that due to the lack of explanation of the flyby anomaly one should use these opportunities in order to carry through a better observation of the Rosetta flybys. A better data basis then will enable one to establish a correlation between the observed velocity increase and the orbital parameters like eccentricity, perihelion distance to the Earth, perihelion velocity, or inclination. In particular, a continuous observation (Doppler tracking, ranging, positioning, and perhaps other data from the spacecraft like temperature, pressure, etc) also should give hints to the particular direction of the local acceleration and also on the strength and, thus, to the position dependence of the anomalous force. Furthermore, it would also be of great importance to know whether after a flyby the direction of the motion of the spacecraft also is different compared with the standard result. These informations are extremely important for the search for a conventional explanation of this effect or for the modeling of a new force. 

As a consequence, a {\it complete, accurate and continuous observation of spacecraft during flybys is very important in order to study the nature of this unexplained velocity increase}. It is an advantage of these planned observations that they will in no way whatsoever modify the mission scenario. 

Furthermore, since new missions like the Mars Reconnaissance Orbiter will measure very precisely the gravitational field of Mars, also the 
\begin{itemize}
\item Mars flyby of Rosetta on 25 February 2007
\end{itemize}
should be observed as precise as possible in order to see whether also at Mars a velocity increase will occur. A Mars flyby would provide an excellent augmentation of the Earth flybys. Since Mars possesses other conditions than the Earth (weaker atmosphere, almost no magnetic field, other gravitational field, lower thermal radiation, etc.) many competing effects and also possible explanation schemes based on the fact that the effect as well as the observation occurs at and on the Earth, can be ruled out. Therefore {\it the effect, if it will be observed also at Mars, then will turn out to be universal and beyond any doubt and will become an extremely important science case}. In the case of Mars it is of course impossible to have a continuous observation (at least due to a Mars eclipse), but the initial and final velocity should be measured with the best possible precision. Furthermore, if this effect really seems to be existent, then a dedicated mission with a good drag--free control system (which can control the acceleration better than $10\;{\rm m/s}^2$ for 1 s measurement time) for well--defined flybys at Earth for small perigees and, thus, for extreme parameter values, might be very helpful for a even better exploration of this effect.

\section{The increase of the Astronomical Unit}

\subsection{The observation}

From the analysis of radiometric measurements of distances between the Earth and the major planets including observations from Martian orbiters and landers from 1961 to 2003 a secular increase of the Astronomical Unit of approximately 10 m/cy has been reported \cite{KrasinskyBrumberg04} (see also the article \cite{Standish05} and the discussion therein). 

\subsection{Search for explanation}

\paragraph{Time--dependent gravitational constant and velocity of light}

This increase cannot be explained by a time--dependent gravitational constant $G$ because the $\dot G/G$ needed is larger than the restrictions obtained from LLR. 

It has also been speculated that a time--dependent change in the velocity of light can be responsible for this effect. Indeed, if the speed of light becomes smaller, than ranging will simulate a drift of distances. However, a inspection of Kepler's third law
\begin{equation}
\frac{T^2}{a^3} = \frac{4\pi^2}{G M_\odot}
\end{equation}
shows that, if one replaces the distance $a$ by a ranging time $a = c t$, then effectively the quotient $G/c^3$ appears. Only this combination of the gravitational constant and the speed of light governs the ratio between the orbit time, in our case the orbit time of the Earth. Consequently, a time--dependent speed of light is equivalent to a time--dependent gravitational constant. Since the latter has been ruled out to be possibly responsible for an increase of the Astronomical Unit, also a time--dependent speed of light has to be ruled out. 

\paragraph{Cosmic expansion}

The influence of cosmic expansion by many orders of magnitude too small, see Sec.\ref{GenCosmExp}. Neither the modification of the gravitational field of the Sun nor the drag of the planetary orbits due to the expansion is big enough to explain this drift. 

\paragraph{Clock drift}

An increase of ranged distances might also be due to a drift of the time scale of the form $t \rightarrow t + \alpha t^2$ for $\alpha > 0$. This is of the same form as the time drift needed to account for the Pioneer anomaly. From Kepler's third law one may ask which $\alpha$ is suitable in order to simulate the increase of the Astronomical Unit. One obtains $\alpha \approx 3 \cdot 10^{-20} \; {\rm s}^{-1}$ what is astonishing close to the clock drift needed for a clock drift simulation of the pioneer anomaly, see Eq.(\ref{PioneerClockDrift}) and below. 

\section{The quadrupole and octupule anomaly}

Recently an anomalous behavior of the low--$l$ contributions to the cosmic microwave background has been reported. It has been shown that (i) there exists an alignment between the quadrupole and octupole with $> 99.87$\% C.L. \cite{OliveiraCostaetal05}, and (ii) that the quadrupole and octupole are aligned to Solar system ecliptic to $> 99$\% C.L. \cite{Schwarzetal04}. No correlation with the galactic plane has been found. 

The reason for this is totally unclear. One may speculate that an unknown gravitational field within the Solar system slightly redirects the incoming cosmic microwave radiation (in the similar way as a motion with a certain velocity with respect to the rest frame of the cosmological background redirects the cosmic background radiation and leads to modifications of the dipole and quadrupole parts). Such a redirection should be more pronounced for low--$l$ components of the radiation. It should be possible to calculate the gravitational field needed for such a redirection and then to compare that with the observational data of the Solar system and the other observed anomalies.

\section{Summary of anomalies}

\subsection{Summary}

In our opinion, these above described anomalies split into three groups according to their observational status: (1) The observations related to dark matter and dark energy are beyond any doubt, (2) the Pioneer anomaly and the flyby anomaly are on a good basis while (3) the increase of he Astronomical Unit as well as the quadrupole and octupole anomaly are still under debate. 

\begin{table}[h]
\caption{List of anomalies and their status}
\begin{center}
\begin{tabular}{|p{4.5cm}|p{4.5cm}|p{4.5cm}|} \hline
{\bf Anomaly} & {\bf Observational status} & {\bf Interpretation} \\ \hline
Dark Energy, \hfil\break Dark Matter & well established 
& 
under discussion \\ \hline
Pioneer Anomaly, \hfil\break Flyby anomaly & quite well established 
& 
unclear \\ \hline
Quadrupole anomaly, \hfil\break increase of AU & 
unclear & 
unclear \\ \hline
\end{tabular}
\end{center}
\end{table}

This list of anomalies and unexplained phenomena immediately induces a bunch of tasks and questions:
\begin{itemize}\itemsep=-2pt
\item For each of these phenomena, except for the Dark Energy and Dark Matter, one still should try to find a systematic cause.
\item One also should try to find conventional explanations of these effects within General Relativity leading perhaps to an effect within conventional physics not considered up to now.
\item A big step in understanding these effects might be to find a relation between two or more of the unexplained phenomena (all of these anomalies very probably are no isolated phenomena).
\item This is tantamount to the important question whether all or at least some of these effects have a common cause.
\item Motivated by the surprising almost perfect equality of $a_{\rm Pioneer} = c H$ one should analyze the influence of the cosmic expansion on the physics within gravitationally bound systems. There already appeared quite some literature related this topic \cite{Krasinski97,Bonnor00,LaemmerzahlDittus05,CarreraGiulini06}, and references therein. 
\item Since there seem to exist some strange phenomena within our Solar system, one should look whether it might be possible to observe similar things in other gravitating systems like binary systems, binary pulsars, stars moving around the black hole in the center of our galaxy, etc.
\item Is something wrong with weak gravity or gravity at large distances?
\item Furthermore, one should propose new dedicated missions and space experiments.
\end{itemize}

\subsection{Other anomalies?}

There is one further observation which status is rather unclear bit which perhaps may fit into the other observations. This is the observation of the return time of comets: Comets usually come back a few days before they are expected when applying ordinary equations of motion. The delay usually is assigned to the outgassing of these objects. In fact, the delay is used for an estimate of the strength of this outgassing. On the other hand, it has been calculated in \cite{PageDixonWallin05} that the assumption that starting with 20 AU there is an additional acceleration of the order of the Pioneer anomaly also leads to the effect that comets come back a few days earlier. It is not clear whether this is a serious indications but a further study of the trajectories of comets certainly is worthwhile. 

\section{Ways to describe the effects}\label{Description}

Many approaches have been attempted to explain these anomalies. In most of these attempts a link between one phenomenon and the issues of (i) the influence of the expansion of the universe on the physics within our Solar system, (ii) dark energy, and (iii) dark matter has been tried to establish. We like to emphasize again that indeed it should be the first thing to explore whether there are links between these various observed and unexplained phenomena. It should be strange if all these unexplained phenomena will be really independent of each other, that is, are not linked by a common (perhaps new) physical principle. 

However, it also seems to be a rather difficult task to find similarities between the Pioneer anomaly and the flyby anomaly. 

\subsection{Cosmological constant}

One first very simple attempt to generalize General Relativity and to incorporate also findings related to dark energy is to describe the physics of gravitating bodies within a theory involving a cosmological constant. The corresponding fields equations are 
\begin{equation}
R_{\mu\nu} - \frac{1}{2} g_{\mu\nu} R + g_{\mu\nu} \Lambda = \kappa T_{\mu\nu}
\end{equation}
where $R_{\mu\nu}$ is the Ricci tensor, $R$ the Ricci scalar, and $T_{\mu\nu}$ the energy momentum tensor. 

In a first step one may use this equation in order to describe the gravitational field of the Sun and to describe the physics within the Solar system (Perihelion shift, light deflection, gravitational redshift, gravitational time delay, geodetic precession) in \cite{KagramanovaKunzLaemmerzahl06} as well as by Iorio \cite{Iorio05b} and Sereno and Jetzer \cite{SerenoJetzer06}. It comes out that the ordinary cosmological constant $\Lambda = \Lambda_0 = 10^{-52}\;{\rm m}^{-2}$ has no observable influence on all the Solar system tests. One also can show that if one assumes a constant $\Lambda$ such large that it may explain the Pioneer acceleration, then this is ruled out by the perihelion shift observation \cite{KagramanovaKunzLaemmerzahl06}. Therefore, the cosmological constant cannot be an explanation for the Pioneer anomaly. In the paper by Jetzer and Sereno \cite{JetzerSereno06} also the influence of the cosmological constant on the motion of binary systems has been evaluated with the result that in the future binary systems may be precise enough to ''see'' the cosmological constant. Some of these effects have also been considered in a Kerr--de Sitter spacetime \cite{KerrHauckMashhoon03}. 

The calculation of Solar system effects in a Schwarschild--de Sitter space--time is certainly only a simple first step. One may redo this kind of calculation in a more general context like quintessence \cite{Wetterich88,PeeblesRatra03} (for a first step in this direction see \cite{Mbelek04}), in varying $G$ scenarios \cite{ReuterWeyer04}, in dilaton scenarios \cite{DamourPiazzaVeneziano02,DamourPiazzaVeneziano02a}, and in braneworld models \cite{DvaliGabadadzePorrati00}. 

\subsection{The influence of the expansion of the universe}\label{GenCosmExp}

We show that the expansion of the universe hat no measurable influence whatsoever on the physics within the Solar system. This includes the modifications of the gravitational field created by the Sun and the planetary orbits.

\paragraph{Modified gravitational field of the Sun}

It is also conceivable that the cosmic expansion may weaken the gravitational field of the Sun. However, it can be shown \cite{LaemmerzahlDittus05} that the corresponding effect is far beyond being observable. Starting from an expansion of the local metric around the cosmological background metric $b_{\mu\nu}$
\begin{equation}
g_{\mu\nu} = b_{\mu\nu} + h_{\mu\nu} \, , \quad \hbox{where} \quad  h_{\mu\nu} \ll b_{\mu\nu} \, .
\end{equation}
we obtain linearized Einstein equations for $h_{\mu\nu}$ \cite{MTW73}
\begin{equation}
g^{\rho\sigma} D_\rho D_\sigma \bar h_{\mu\nu} + 2 g^{\rho\sigma} R^\kappa{}_{\mu\rho\nu} \bar h_{\kappa\sigma} = 16 \pi G T_{\mu\nu} \, ,
\end{equation}
The static solution for a small spherically symmetric mass distribution is given by
\begin{equation}
h_{00} = \frac{2 G M}{R} \frac{\cos\bigl(\sqrt{6} |\dot R| r\bigr)}{r} = \frac{G M}{R r} \left(1 - 3 H^2 (R r)^2 \pm \ldots\right) 
\end{equation}
At lowest order we obtain the standard Newtonian potential with the measured distance $R(t) r$. Since the modification is quadratic in the Hubble constant, the Newtonian potential practically does not participate in the cosmic expansion. This confirms the findings in \cite{KrasinskyBrumberg04}. 

\paragraph{Modified planetary orbits}

If the planetary orbits may expand due to a drag from the cosmic expansion, then this might be interpreted as if the length unit, Astronomical Unit, increases so that it might appear that other distances may become smaller. However, from the theorem of adiabatic invariance \cite{LandauLifshitz76} it can be shown that the parameters of the planetary orbits remain the same with very high stability \cite{LaemmerzahlDittus05}. This stability is characterized by the factor $\exp(-t/T)$ where $t$ is the time scale of the planetary orbit and $T$ the time scale of the temporal change of the environment, in our case the Hubble time. Therefore, $t/T$ is of the order $10^{10}$ so that the planetary orbits are extremely stable.   

Another approach is based on the geodesic deviation which relates the relative acceleration of freely moving particles with the curvature of space--time, in our case with the expansion of the universe. This has been taken in \cite{Cooperstocketal98} as the basis for estimates that this deviation is much too small to be of any influence in the Solar System or even within galaxies. For a broader discussion of this approach, see the recent valuable review of Carrera and Giulini \cite{CarreraGiulini06}. 

\subsection{General approach to describe a modified particle dynamics}

Gravity describes or, equivalently, can be characterized by the behavior of light and point particles \cite{EhlersPiraniSchild72}. Light is related to a space--time metric while point particles are related to a geodesic equation. In Riemannian geometry, the mathematical model for General Relativity, the equation of motion for particles and the behavior of light are related: The equation of motion for particles is completely determined by the space--time metric. For general analysis of anomalous gravitational effects it therefore may be appropriate to start a by a general ansatz for the metric and of the equation of motion for particles\footnote{In the axiomatic approach \cite{EhlersPiraniSchild72} the mathematical relation between the point particle dynamics and the properties of light came in through a compatibility condition. One feature of this condition is a causal dynamics of point particles; particles are not allowed to move faster than light. The requirement of this condition implies a rather restricted geometrical structure, namely a Weylian structure. Since in \cite{EhlersPiraniSchild72} this condition has been applied to autoparallel curves and since we are here more general in admitting also non--linear connections as they appear, e.g., in a Finslerian context, the resulting allowed particle dynamics are certainly still more general than a Weylian structure. However, this has to be worked out explicitly.}. 

In a weak field approximation the metric can be written in the form 
\begin{equation}
g_{\mu\nu} = \begin{pmatrix} 1 - U & g_{0i} \\ g_{i0} & \delta_{ij} \left(1 - V\right) \end{pmatrix}
\end{equation}
where $U$, $V$, and $g_{0i} = g_{i0}$ are assumed to be small quantities. $U$ may be identified with the Newtonian potential. We denote $ds^2 = g_{\mu\nu} dx^\mu dx^\nu$. 

By definition, the readout of moving clocks is given by 
\begin{equation}
T = \int_{\rm worldline\; of\; clock} ds \; \approx \int \bigl(1 - U + \dot x^2 + V \dot x^2\bigr) dt + \int \mbox{\boldmath$h$} \cdot d\mbox{\boldmath$x$} 
\end{equation}
which is the proper time of the clock. A clock at rest will measure the gravitational redshift given by 
\begin{equation}
\frac{\nu_1}{\nu_2} = \sqrt{\frac{g_{00}(x_1)}{g_{00}(x_2)}} \approx 1 + U(x_2) - U(x_1) \, .
\end{equation}

Now we are going to give a general description of the equation of motion for particles. This approach is according to the philosophy of axiomatic approaches to theories of gravity, see, e.g., \cite{EhlersPiraniSchild72,Laemmerzahl96}. The equation of motion for a particle respecting the Universality of Free Fall can be represented as 
\begin{equation}
0 = v^\nu \partial_\mu v^\nu = H^\mu(x, v) = \left\{\begin{smallmatrix}\mu \\ \rho\sigma\end{smallmatrix}\right\} v^\mu v^\sigma + \gamma^\mu(x, v)
\end{equation}
where $\left\{\begin{smallmatrix}\mu \\ \rho\sigma\end{smallmatrix}\right\}$ is the Christoffel connection and $\gamma^\mu(x, v)$ some vector valued function of the position and the velocity. Since no particle parameter enters this equation, the Universality of Free Fall automatically is preserved. 

From this general equation of motion, we can derive the 3--acceleration
\begin{eqnarray}
\frac{d^2 x^i}{dt^2} & = & - \left(\left\{\begin{smallmatrix}i \\ \mu\nu\end{smallmatrix}\right\} - \left\{\begin{smallmatrix}0 \\ \mu\nu\end{smallmatrix}\right\} \frac{dx^i}{dt}\right) \frac{dx^\mu}{dt} \frac{dx^\nu}{dt} + \frac{1}{\left(\frac{dt}{ds}\right)^2} \left(\gamma^i(v, x) - \frac{dx^i}{dt} \gamma^0(v, x)\right) \nonumber\\
& = & - \left(\left\{\begin{smallmatrix}i \\ 00\end{smallmatrix}\right\} - \left\{\begin{smallmatrix}0 \\ 00\end{smallmatrix}\right\} \frac{dx^i}{dt}\right) - 2 \left(\left\{\begin{smallmatrix}i \\ j 0\end{smallmatrix}\right\} - \left\{\begin{smallmatrix}0 \\ j 0\end{smallmatrix}\right\} \frac{dx^i}{dt}\right) \frac{dx^j}{dt} - \left(\left\{\begin{smallmatrix}i \\ jk\end{smallmatrix}\right\} - \left\{\begin{smallmatrix}0 \\ jk\end{smallmatrix}\right\} \frac{dx^i}{dt}\right) \frac{dx^j}{dt} \frac{dx^k}{dt} \nonumber\\
& &  + \frac{1}{\left(\frac{dt}{ds}\right)^2} \left(\gamma^i(v, x) - \frac{dx^i}{dt} \gamma^0(v, x)\right) \nonumber\\
& \approx & \underbrace{\partial_i U}_{\rm Newton} + \underbrace{(\partial_i h_j - \partial_j h_i) \dot x^j}_{\rm Lense-Thirring} + \dot x^2 \partial_i V + \dot x^i \dot V + \Upsilon^i + \Upsilon^i_j \dot x^j + \Upsilon^i_{jk} \dot x^j \dot x^k + \ldots \, , \label{GenEqMotion}
\end{eqnarray}
where we neglected all relativistic corrections since these play no role in the Pioneer and flyby anomalies. Note also that the Universality of Free Fall is respected. However, it is no longer possible to make a  transformation to a coordinate system so that gravity disappears at one point (Einstin's elevator is not possible). Physically this means that, e.g., the acceleration of a body toward the Earth can depend on the velocity of the body: Differently moving bodies feel a different gravitational acceleration which, however, does not depend on the composition or the weight of the (point--like) body. 

The first term in (\ref{GenEqMotion}) is the ordinary Newtonian acceleration and the second term the action of the gravitomagnetic field on the orbit of a satellite which has been observed by LAGEOS with a 10\% accuracy \cite{Ciufolini04}. This field also acts on spinning objects like gyroscopes and should be confirmed by GP-B with an accuracy better than 1\%. The other terms are hypothetical terms beyond ordinary post--Newtonian approximation. 

The $V$ term which can be motivated by a running coupling constant to be proportional to the distance, $V \sim r^2$ has been introduced by Jaeckel and Reynaud \cite{JaekelReynaud05} in order to describe the constant anomalous Pioneer acceleration. The other terms, most of then are velocity dependent, have not yet been analyzed. The influence of an arbitrary force on the trajectories of planets has been analyzed recently in \cite{Iorio06} with the main conclusion that any radial force which might be considered as being responsible for the Pioneer anomaly is not compatible with the recent analysis of the motion of the outer planets. This indicates that the modification of the equation of motion should include velocity--dependent terms. 

The coefficients $\Upsilon^i_{jk\ldots}$ depend on the position only and may vanish for vanishing gravitating mass. Therefore, the coefficients can contain $M$, $r$, $r^i$, and $\partial_i$ only. Accordingly, these coefficients can be of the form
\begin{eqnarray}
\Upsilon^i & = & A_{11} \frac{G M}{r^2} \frac{r^i}{r} \\
\Upsilon^i_j & = & A_{21} \frac{G M}{r^2} \frac{r^i r^j}{r^2} + A_{22} \frac{G M}{r^2} \delta^i_j \\
\Upsilon^i_{jk} & = & A_{31} \frac{G M}{r^2} \frac{r^i r^j r^k}{r^3} + A_{32} \frac{G M}{r^2} \frac{r^i}{r} \delta_{jk} + A_{33} \frac{G M}{r^2} \frac{r^j}{r} \delta^i_k \, .
\end{eqnarray}
Here it is understood that the influence of the gravitating body, that is, all the $\Upsilon$--coefficients, vanishes at spatial infinity. This is certainly true for a description of the flyby anomaly but may be relaxed for the Pioneer anomaly. Terms which do not vanish at spatial infinity but are of Newtonian form at small distances are polynomials $r^l$, $l \geq 1$.

The above terms lead to accelerations
\begin{eqnarray}
\ddot x^i & = & A_{11} \frac{G M}{r^2} \frac{r^i}{r} \\
\ddot x^i & = & A_{21} \frac{G M}{r^2} \frac{r^i \mbox{\boldmath$r$} \cdot \dot{\mbox{\boldmath$r$}}}{c r^2} + A_{22} \frac{G M}{r^2} \frac{\dot r^i}{c} \nonumber\\
& = & (A_{21} + A_{22}) \frac{G M}{r^2} \frac{r^i \mbox{\boldmath$r$} \cdot \dot{\mbox{\boldmath$r$}}}{c r^2} + A_{22} \frac{G M}{r^2} \frac{\dot r^i_\bot}{c} \label{Upsilondotxterm} \\
\ddot x^i & = & A_{31} \frac{G M}{r^2 c^2} \frac{r^i (\mbox{\boldmath$r$} \cdot \dot{\mbox{\boldmath$r$}})^2}{r^3} + A_{32} \frac{G M}{c^2 r^2} \frac{r^i}{r} \dot r^2 + A_{33} \frac{G M}{c^2 r^2} \frac{\dot r^i}{r} (\mbox{\boldmath$r$} \cdot \dot{\mbox{\boldmath$r$}}) \, ,
\end{eqnarray}
where $r^i_\bot = r^i - r^i (\mbox{\boldmath$r$} \cdot \dot{\mbox{\boldmath$r$}})/r^2$ is the component of the body's velocity orthogonal to the connecting vector $\mbox{\boldmath$r$}$, and the $A_{ij}$ are some numerical factors.

The first term associated with $A_{11}$ is of Newtonian form and can be combined with the already existing one which amounts to a redefinition of the gravitational constant. The $A_{22}$ term describes an additional acceleration in direction of the velocity. It fades away for large $r$. The $A_{21}$ term projects the component of the velocity which is parallel to the connecting vector and leads to an acceleration in direction of the connecting vector. This term vanishes at the perigee. 

These $A_{21}$-- and $A_{22}$--terms may be chosen in such a way that they have the potential to describe an increase of the velocity during a flyby: Near the Earth the $A_{22}$--term is dominant since there the connecting vector is more or less orthogonal to the velocity vector. For large $r$ both terms contribute. That is
\begin{equation}
\ddot x^i = \begin{cases} A_{22} \dfrac{G M}{r^2} \dfrac{\dot r^i}{c} & \hbox{for}\; r \approx r_{\rm perigee} \\  (A_{21} + A_{22}) \dfrac{G M}{r^2} \dfrac{\dot r^i}{c} & \hbox{for} \; r \; \hbox{large} \end{cases}
\end{equation} 
Therefore, in principle it is possible to have an acceleration near the perigee (for $A_{22} > 0$) and a deceleration for large distances (for $A_{21} + A_{22} < 0$). Note that for a typical perigee and velocity at perigee the acceleration at perigee for $A_{22} = 1$ is about $10^{-4}\;{\rm m/s}^2$ what is just the value given for a typical Earth flyby, see Sec.\ref{flybyaccel}. However, this model does not include the Pioneer deceleration because the acceleration is not constant for large $r$. What is needed now is a general discussion of the influence of a term of the form (\ref{Upsilondotxterm}) on general features planetary and satellite orbits, e.g., the perihelion shift, and to compare this with observations. Corresponding work is in progress.  

The $A_{32}$ term just adds to the $\dot x^2 \partial_i V$ term. The $A_{31}$ and the $A_{33}$ term both first project the velocity in direction of the connecting vector and then make out of this an acceleration in direction of the connecting vector and in direction of the velocity. These $A_{3i}$--terms are about 5 orders of magnitude smaller than the $A_{2i}$--terms and can, thus, play no role in an explanation of the flyby anomaly. The higher order terms will be of more complicated but similar structure.  

In general, this equation of motion does not respect energy conservation: multiplication of (\ref{GenEqMotion}) with the velocity yields
\begin{eqnarray}
\frac{d}{dt} \left(\tfrac{1}{2} \dot{\mbox{\boldmath$x$}}^2 - U\right) & = & 2 \dot x^2 \dot V + \dot{\mbox{\boldmath$x$}} \cdot \mbox{\boldmath$\Upsilon$} + \Upsilon^i_j \dot x^j \dot x^i + \Upsilon^i_{jk} \dot x^j \dot x^k \dot x^i + \ldots
\end{eqnarray}
Therefore, the terms on the right hand side might be candidates for effects reducing or enlarging the kinetic energy of moving bodies and, thus, may play a role in the description of the flyby or the Pioneer anomaly. 

It should be clear from the independence of the metric from the equation of motion for point particles, that it is necessary both to track position and velocity of the satellite and to have a clock on board in order to determine all components of the space--time metric. 

\section{Summary and outlook}

We collected the anomalies related to the physics of the Solar system and discussed to some extend the error sources and possibilities to explain these anomalies. In particular, we tried to find similarities or fundamental differences between these anomalies. 

As final statement we like to stress that there are at least three important science cases related to the exploration of these anomalies which we strongly suggest to be tackled in the near future:
\begin{enumerate}\itemsep=-2pt
\item Analysis of the complete set of Pioneer data.
\item Continuous and complete (velocity, distance, time on board, and direction) observations of future flybys.
\item Search for clock drifts by comparison of clock rates on Earth with clocks defined by astrophysical systems. 
\end{enumerate} 

\section*{Acknowledgement}

We like to thank J. Anderson, D. Giulini, S. Grotian, P. Jetzer, G. Krasinsky, B. Mashhoon, T. Morley, E. Pitjeva, G. Sch\"afer, S. Solanki, and S. Turyshev for support, informations, and various discussions. Financial support from the German Research Foundation DFG and the German Aerospace Agency DLR is gratefully acknowledged. 


\end{document}